\documentclass[twocolumn,amsmath,trackchanges]{aastex702}

\begin{document}

\title{A Template-Based Search for Large-Scale-Structure--Correlated Anisotropy in the Nanohertz Gravitational-Wave Background Using the Public NANOGrav 15-Year Data Set}


\author[orcid=0000-0003-0065-8622]{Yun Fang}
\affiliation{Institute of Fundamental Physics and Quantum Technology, Ningbo University, Ningbo, 315211, China}
\affiliation{BINGO Center, Ningbo University, Ningbo, 315211, China}
\affiliation{School of Physical Science and Technology, Ningbo University, Ningbo, 315211, China}
\email[show]{fangyun@nbu.edu.cn}

\begin{abstract}

	Recent PTA analyses reporting evidence for a nanohertz common-spectrum process motivate targeted tests of whether any anisotropic component of the stochastic gravitational-wave background (SGWB) is correlated with the nearby large-scale structure (LSS), as anticipated for an astrophysical background dominated by supermassive black hole binaries. We present the first Bayesian PTA likelihood analysis that embeds an externally observed, full-sky galaxy-survey LSS template directly as an overlap-reduction-function (ORF) component. Using the 2MASS Photometric Redshift (2MPZ) galaxy catalog, we construct low-multipole LSS--correlated ORF templates in two redshift slices ($0<z\le0.1$ and $0.1<z\le0.2$) and model PTA cross-correlations as $\Gamma_{ab}=\Gamma^{\rm HD}_{ab}+\sum_i \epsilon_i\,\Gamma^{\rm LSS(i)}_{ab}$, 
where $\epsilon_i$ quantifies the amplitude of an SGWB component whose angular correlations project onto the fixed 2MPZ LSS templates. Applying this framework to the NANOGrav 15-year dataset, we find no evidence for an LSS-correlated component: the inferred $\epsilon_i$ values are consistent with zero in both single-bin and two-bin analyses (e.g., $\epsilon_1=0.20^{+1.68}_{-1.66}$ and $\epsilon_2=-0.11^{+2.04}_{-1.83}$; 68\% credible intervals), and Bayes factors favor the isotropic Hellings--Downs hypothesis ($\mathcal{B}_{{\rm HD+LSS}_1,{\rm HD}}=0.40$, $\mathcal{B}_{{\rm HD+LSS}_2,{\rm HD}}=0.43$, $\mathcal{B}_{{\rm HD+LSS}_{1+2},{\rm HD}}=0.11$). We therefore place upper limits on any 2MPZ-traced, LSS-correlated contribution to the SGWB at $z<0.2$. More broadly, our framework provides a reproducible pathway for incorporating observed LSS information into PTA anisotropy searches and naturally motivates extensions to finer redshift tomography and next-generation PTA datasets.

\end{abstract}


\section{Introduction}\label{sec:intro}

Pulsar timing arrays (PTAs) use precision timing of an ensemble of millisecond pulsars to probe gravitational waves (GWs) in the nanohertz band, enabling searches for both a stochastic gravitational-wave background (SGWB) and individually resolvable sources \citep{1978SvA....22...36S,1979ApJ...234.1100D,1990ApJ...361..300F,2011gwpa.book.....C,maggiore2008gravitational}.
A global PTA effort is now in place, including the Parkes PTA (PPTA)\citep{manchester2013pulsar,2023ApJ...951L...6R}, the 
European PTA (EPTA)\citep{2013CQGra..30v4009K,2023A&A...678A..50E}, the North American Nanohertz 
Observatory for Gravitational Waves (NANOGrav)\citep{2013CQGra..30v4008M,NANOGrav_2023_SGWB_evidence}, the Indian PTA (InPTA)\citep{2018JApA...39...51J,2023A&A...678A..50E}, the Chinese PTA (CPTA)\citep{2011IJMPD..20..989N,2009A&A...505..919S,2023RAA....23g5024X}, and the MeerKAT PTA (MPTA)\citep{2023MNRAS.519.3976M}.
By combining data and expertise through the International PTA (IPTA)\citep{2013CQGra..30v4010M,10.1093/mnras/stw347,2019MNRAS.490.4666P} and looking ahead to next-generation facilities such as the Square Kilometre Array \citep{Lazio2013SKA,2017PhRvL.118o1104W}, PTAs are entering an era in which the SGWB is becoming accessible to detailed characterization \citep{NANOGrav_2023_SGWB_evidence,2023A&A...678A..50E,NANOGrav_2023_Anisotropy}.
Recent analyses by CPTA, NANOGrav, EPTA(+InPTA), and PPTA report evidence for a nanohertz common-spectrum process with spatial correlations 
consistent with a GW background \citep{2023RAA....23g5024X,NANOGrav_2023_SGWB_evidence,2023ApJ...951L...6R,2023A&A...678A..50E}.

With accumulating evidence for a nanohertz common-spectrum process, a central question is its physical origin.
Different source classes predict distinct spatial statistics for the SGWB, making anisotropy a powerful complement to spectral characterization \citep[e.g.,][]{Taylor_Gair_2013,Cornish_2014,Mingarelli_2017_LocalLandscape,Hotinli_2019,NANOGrav_2023_Anisotropy}.
The interpretation most often discussed is an SGWB produced by the cosmic population of inspiralling supermassive black hole binaries (SMBHBs) \citep[e.g.,][]{Phinney_2001, Jaffe_Backer_2003,  sesana2013gravitational, 2023ApJ...952L..37A, Bi:2023tib}, while alternative early-Universe mechanisms (e.g., cosmic defects or phase transitions) remain viable given current uncertainties \citep[e.g.,][]{PhysRevLett.69.2026,Caprini_2018, PhysRevLett.126.041304, NANOGrav23_new_physics}.
To distinguish these possibilities, it is essential to go beyond an isotropic characterization and test whether the SGWB exhibits spatial statistics that are characteristic of its origin.

Anisotropy and its correlation with the large-scale structure (LSS) provide a particularly direct discriminant.
If a significant fraction of the PTA-band SGWB is sourced by SMBHBs, departures from statistical isotropy are generically expected because the signal is weighted toward low redshifts where galaxies and dark-matter halos are clustered.
This links the SGWB anisotropy to the cosmic web: SMBHs co-evolve with their host galaxies and trace the assembly history of dark-matter halos, as suggested by SMBH--galaxy and SMBH--halo scaling relations \citep[e.g.,][]{Ferrarese_2000,Gebhardt:2000fk,Kormendy:2013dxa,Reines_2015,JWST_2025_SMBH_halo}, and AGNs preferentially reside in massive halos and overdense environments \citep[e.g.,][]{Wake_2008_radioAGN, Donoso_2010, Hickox_2011, Hashiguchi_2023}.
In addition, because the nanohertz SGWB is built from a finite number of binaries, Poisson fluctuations in the number and sky distribution of contributing systems generate a ``shot-noise'' anisotropy component \citep[e.g.,][]{Ravi_2012,Cornish_2013, Allen2024_PTA_anisotropy}, superposed on any LSS-driven modulation.
By contrast, many early-Universe mechanisms predict SGWBs that are statistically isotropic at leading order, with only small departures from isotropy expected \citep[e.g.,][]{Guzzetti_2016,Taylor_Gair_2013,Alba_Maldacena_2016,Caprini_2018,Bartolo_2020}.
Characterizing SGWB anisotropy therefore provides a complementary probe for distinguishing astrophysical and cosmological interpretations of the signal \citep[e.g.,][]{Mingarelli_2013,Taylor_Gair_2013,Cornish_2014,Hotinli_2019,NANOGrav_2023_Anisotropy}.

A substantial literature has developed the theoretical and data-analysis foundations for PTA anisotropy.
Early work introduced multipolar descriptions of the GW power distribution on the sky and generalized overlap-reduction functions (ORFs) for anisotropic PTA cross-correlations \citep{Mingarelli_2013,Gair_2014_prd}, and developed Bayesian search pipelines to infer angular structure in simulated PTA data \citep{Taylor_Gair_2013}. 
Dedicated searches on real PTA data have reconstructed the GW sky using model-independent spherical-harmonic and map-based methods and, despite occasional low-significance hotspots, have found no statistically significant anisotropy beyond the monopole, placing upper limits on low-$\ell$ power \citep[e.g.,][]{Taylor_2015_PRL,NANOGrav_2023_Anisotropy,MPTA_2024_Anisotropy,PPTA_2026_Anisotropy}.
A practical complication is that realistic nanohertz SGWBs are subject to large realization scatter (``cosmic variance") from a finite population of SMBHBs, in which a few bright binaries can produce a ``bumpy" GW sky and potentially mask more specific clustering signatures \citep[e.g.,][]{Taylor_2020PRD,Becsy_2022,Konstandin_2024}.
At the same time, recent methodological work emphasizes that sensitivity can improve when information from multiple anisotropy modes is combined optimally, motivating low-dimensional, physically grounded anisotropy models in the low-SNR regime \citep[e.g.,][]{Hotinli_2019,Yacine_2021}. 
In parallel, local-Universe galaxy catalogs and cosmological simulations have been used to predict anisotropic nanohertz GW ``landscapes'' and source clustering \citep[e.g.,][]{Mingarelli_2017_LocalLandscape, Sah_2024MN, Yang_2025PTA}, and to incorporate external astrophysical priors in targeted PTA searches \citep[e.g.,][]{Arzoumanian_2021}.
Recent work further shows that cross-correlations with galaxy maps can isolate the component that traces the underlying LSS and substantially improve detectability even in the presence of strong Poisson shot noise \citep[e.g.,][]{Semenzato2024,Cusin_2025_CrossCorre}. 

In this work, we adopt a complementary, template-based strategy for testing SGWB anisotropy. We use the observed galaxy overdensity field to construct redshift-sliced, LSS-derived ORF templates, and model the PTA cross-correlation as a linear superposition of the isotropic HD form and an LSS-tracing contribution with mixing coefficients that are inferred directly from the NANOGrav 15 yr data \citep{NANOGrav_2023_SGWB_evidence,NANOGrav_dataset_2023}. This framework enables an explicit test of the hypothesis that any anisotropic component of the PTA-band SGWB follows the angular distribution of matter in the nearby Universe. Concretely, we employ the 2MASS Photometric Redshift (2MPZ) catalog \citep{Bilicki20142MPZ} to build the LSS templates. While current Bayes factors prefer the isotropic model, the data nonetheless place upper limits on the LSS-correlated components. As a conservative first step, we restrict attention to $z<0.2$ and relatively narrow redshift slices, enabling controlled validation of the LSS reconstruction and the resulting ORF templates. This low-redshift restriction is a conservative, survey-limited first step, and our results should therefore be interpreted as constraints on the 2MPZ-traced, low-redshift LSS-correlated component of the SGWB anisotropy, rather than on the full SMBHB-induced anisotropy budget.

This paper is organized as follows. In Section~\ref{sec:definition_ORF} we review the basic PTA response to an isotropic and anisotropic SGWB and define the corresponding ORFs. In Section~\ref{sec:lss_motivation_eps_c} we motivate an LSS-informed anisotropy ansatz and derive the LSS-template ORF parameterization, leading to the binned mixed-ORF model. In Section~\ref{sec:lss_alm} we describe the construction of redshift-sliced 2MPZ overdensity maps and our cut-sky MAP reconstruction of low-$\ell$ spherical-harmonic coefficients used to build the LSS ORF templates. In Section~\ref{sec:results_lss_orf} we present Bayesian constraints from the NANOGrav 15-year dataset on single-bin and two-bin LSS-modulated ORF models and report hypermodel Bayes-factor comparisons against the HD-only hypothesis. We discuss the implications, limitations, and prospects for future PTA datasets in Section~\ref{sec:discu_and_concl}. 
Additional diagnostics of the low-$\ell$ reconstruction and the stability of the harmonic truncation are provided in Appendix~\ref{app:lowell_diagnostics}.

\section{Basic concepts of isotropic and anisotropic ORF}
\label{sec:definition_ORF}
In the weak-field, transverse-traceless (TT) gauge, the dimensionless metric perturbation can be expanded as
\begin{align}
	h_{ij}(t,\mathbf{x})
	&=\sum_{A=+,\times}\int \mathrm{d} f\int \mathrm{d} \mathbf{n}\;
	h_A(f,\mathbf{n})\,e^{2\pi i f\left(t-\mathbf{n}\cdot \mathbf{x}\right)}\,
	e^{A}_{ij}(\mathbf{n})\,,
	\label{eq:plane}
\end{align}
where $e^{A}_{ij}$ is the polarization tensor and $h_A(f,\mathbf{n})$ is the amplitude of the GW propagating in direction $\mathbf{n}$ with frequency $f$.
The pulsar redshift $z_a$, for pulsar $a$ located in direction $\mathbf{p}_a$, is given by \citep[e.g.,][]{1979ApJ...234.1100D,2009PhRvD..79h4030A,Gair_2014_prd}
\begin{align}
	z_a(t)
	=&\sum_A\int \mathrm{d} f\int \mathrm{d} \mathbf{n}\; F_a^A(\mathbf{n})\,h_A(f,\mathbf{n})\,
	e^{2\pi i f t}
	\nonumber\\
	&\times \Bigl[1-e^{-2\pi i f L_a(1+\mathbf{n} \cdot \mathbf{ p}_a)}\Bigr].
	\label{eq:z_full}
\end{align}
where $L_a$ is the distance from this pulsar to Earth, and ${F_a^A}(\mathbf{n})$ is the PTA antenna response defined as 
\begin{align}
	F_a^A(\mathbf{n})
	\equiv {p_a^i p_a^j e^A_{ij}(\mathbf{n}) \over 2(1+\mathbf{n}\cdot \mathbf{p}_a)}.
	\label{eq:FA}
\end{align}

The timing residual is the time integral of the redshift:
\begin{align}
	r_a(t)=\int^t\!dt'\, z_a(t'). 
\end{align}
Hence, in the frequency domain,
\begin{align}
	\tilde r_a(f) &= \frac{\tilde z_a(f)}{2\pi i f} \nonumber\\
	&=\sum_A\int\! \mathrm{d} \mathbf{n}\; 
	\frac{F_a^A(\mathbf{n})}{2\pi i f}\,h_A(f,\mathbf{n})
	\Bigl[1-e^{2\pi i f L_a(1+\mathbf{n}\cdot \mathbf{p}_a)}\Bigr].
	\label{eq:r_full}
\end{align}

\subsection{ Isotropic background and HD ORF}
\label{subsec:HD_ORF}
Consider two pulsars with residual $\tilde r_a(f)$ and $\tilde r_b(f)$. For an isotropic, stationary, unpolarized stochastic background with one-sided strain PSD $S_h(f)$, we have
\begin{align}
	\big\langle h_A(f,\mathbf{n}) h^*_{A'}(f',\mathbf{n}')\big\rangle
	=\frac{1}{2}\,\delta_{AA'}\,\delta(f-f')\,\frac{\delta^2(\mathbf{n},\mathbf{n}')}{4\pi}\,S_h(f).
	\label{eq:stat}
\end{align}
Here and below, an asterisk denotes complex conjugation of the corresponding Fourier-domain quantity.
Using \eqref{eq:r_full} and \eqref{eq:stat}, the cross-spectrum of the residuals is then
\begin{align}
	\big\langle \tilde r_a(f)\tilde r_b^*(f')\big\rangle
	=\frac{1}{2}\delta(f-f')\, \Gamma_{ab}(f) \frac{S_h(f)}{12\pi^2 f^2}\,,
\end{align}
where 
\begin{align}
	\Gamma_{ab}(f)=\frac{3}{2}
	\sum_{A}\int\frac{\mathrm{d}\mathbf{n}}{4\pi}\;
	F_a^A(\mathbf{n})F_b^A(\mathbf{n})\,\kappa_{ab}(f,\mathbf{n})\,,
	\label{eq:Gamma_full}
\end{align}
defines the frequency-dependent ORF, 
and 
\begin{align}
	\kappa_{ab}(f,\mathbf{n})
	=\Bigl[1-e^{2\pi i f {L_a}(1+\mathbf{n} \cdot \mathbf{p}_a)}\Bigr]
	\Bigl[1-e^{-2\pi i f {L_b}(1+\mathbf{n} \cdot \mathbf{p}_b)}\Bigr].
	\label{eq:kappa}
\end{align}
Under the approximation $2\pi f L\gg 1$, only the constant term (the EE term) survives: $\kappa_{ab}(f,\mathbf{n})\to 1$, 
while the other terms oscillate rapidly and average to zero. 

Using the polarization-sum identity, the angular integration on the sphere yields the standard Hellings--Downs (HD) curve, with the convention $\Gamma_{aa}=1$:
\begin{align}
	\Gamma_{ab}^{\rm HD}(\zeta)
	&= \frac{3}{2}\left[x\ln x-\frac{1}{6}x+\frac{1}{3}\right]\,,
	\nonumber\\
	\label{eq:HD}
\end{align}
where $x \equiv \frac{1-\cos\zeta}{2}$, and $\zeta$ is the separation angle between two pulsars $a$ and $b$.

\subsection{Anisotropic backgrounds and directional ORFs}
\label{subsec:anisotropic_ORF}
If the gravitational wave background has an angular distribution $P(\mathbf{n})$, then
\begin{align}
	\! \big\langle h_A(f,\mathbf{n}) h^*_{A'}(f',\mathbf{n}')\big\rangle
	\!\!=\!\!\frac{1}{2}\,\delta_{AA'}\,\delta(f\!-\!f')\,\delta^2(\mathbf{n},\mathbf{n}')\,S_h(f)\,P(\mathbf{n}),
\end{align}
where $P(\mathbf{n})$ is the angular distribution of GW power on the sky, 
and the ORF becomes
\begin{align}
	\Gamma_{ab}(f)=\frac{3}{2}\sum_A\int \mathrm{d}\mathbf{n}\;
	F_a^A(\mathbf{n})F_b^A(\mathbf{n})\,
	\kappa_{ab}(f,\mathbf{n})\,P(\mathbf{n})\,.
	\label{anisotropy_G_ab}
\end{align}
The angular distribution $P(\mathbf{n})$ could be further expanded in spherical harmonics as $P(\mathbf{n})=\sum_{\ell m}c_{\ell m}Y_{\ell m}(\mathbf{n})$. Take it back into Eq.~(\ref{anisotropy_G_ab}) we have 
\begin{align}
	\Gamma_{ab}(f)&= {3\over 2}
	\sum_{\ell m} c_{\ell m}\,\Gamma_{ab}^{\ell m}(f), 
\end{align}
where
\begin{align}
	\Gamma_{ab}^{\ell m}(f)& =
	\sum_A\int \mathrm{d} \mathbf{n}\;
	F_a^A(\mathbf{n})F_b^A(\mathbf{n})\,\kappa_{ab}(f,\mathbf{n})\,Y_{\ell m}(\mathbf{n})\,.
\end{align}

In the short-wavelength approximation, only the Earth-Earth term survives the integration in Eq.~(\ref{anisotropy_G_ab}), giving 
\begin{subequations}\begin{align}
		&\Gamma_{ab}
		={3\over 2} \int \mathrm{d} \mathbf{n}\; P(\mathbf n)\,\mathcal{F}_{ab}(\mathbf n), \\
		&\mathcal{F}_{ab}(\mathbf n)\equiv 
		\sum_{A=+,\times}F_a^A(\mathbf n)\,F_b^A(\mathbf n)\,.
	\end{align}\label{eq:Gamma_of_P_main}\end{subequations}
The isotropic case corresponds to the single $\ell=0,m=0$ mode, which has $P(\mathbf{n})=1/(4\pi)$ and one recovers the HD correlation pattern.

\section{Relating SGWB anisotropy to clustering of LSS}
\label{sec:lss_motivation_eps_c}

Physically, the nanohertz SGWB is expected to be dominated by a finite population of astrophysical sources, most notably SMBHBs. Since SMBHs typically reside in galactic nuclei and galaxies trace the distribution of dark matter halos \citep[e.g.,][]{Ferrarese_2000,Gebhardt:2000fk,Kormendy:2013dxa,Reines_2015,JWST_2025_SMBH_halo}, SMBHBs are expected to follow the cosmic LSS to some degree. 
This physical picture is also supported by previous studies of astrophysical SGWB anisotropies and their cross-correlations with galaxy/LSS tracers
\citep[e.g.,][]{Cusin:2017fwz,CanasHerrera2020,Yang2023,Semenzato2024}, which suggest that the unresolved SMBHB background is expected to be statistically correlated with the angular distribution of LSS.
Motivated by these considerations, we adopt, as a first step, a phenomenological template-based approach in which the LSS-correlated component of the SGWB anisotropy is taken to be maximally correlated with the observed LSS in each redshift slice:
\begin{align}
	\delta_{\rm gw}(\mathbf{n},z)
	=
	c(z)\,\delta_{\rm gal}(\mathbf{n},z),
	\label{eq:delta_gw_equals_c_delta_gal_z}
\end{align}
where $\delta_{\rm gal}(\mathbf{n},z)$ denotes the galaxy overdensity as a function of sky direction $\mathbf{n}$ and redshift $z$ (e.g., in tomographic redshift slices), and $c(z)$ is a bias-like amplitude that controls the strength of the LSS-tracing component. Under this ansatz, the GW and galaxy fields have identical angular phases within each redshift slice.
A closely related galaxy-overdensity--modulated parametrization of the directional SGWB spectral density can be found in \cite{Grimm_2025_theory,Grimm_2025_numerical}.
We emphasize that this ansatz is not intended to provide a complete description of the true SGWB anisotropy, but rather to define a LSS-correlated ORF template whose amplitude can be directly tested with PTA data.

\subsection{LSS-template ORF}
\label{subsec:lss_orf}

To connect this to the ORF, we allow the angular weight to receive contributions from all redshifts along the line of sight,
\begin{align}
	P(\mathbf{n})
	&=\frac{1}{4\pi}\left[1+\int_{z_{\min}}^{z_{\max}} \mathrm{d}z\; \delta_{\rm gw}(\mathbf{n},z)\right]
	\nonumber\\
	&=\frac{1}{4\pi}\left[1+\int_{z_{\min}}^{z_{\max}} \mathrm{d} z\; c(z)\,\delta_{\rm gal}(\mathbf{n},z)\right].
	\label{eq:P_1_plus_delta_main_z}
\end{align}
Inserting \eqref{eq:P_1_plus_delta_main_z} into \eqref{eq:Gamma_of_P_main} and using linearity yields a decomposition into an
isotropic (HD-like) piece plus an LSS-weighted anisotropic contribution integrated over redshift:
\begin{align}
	\Gamma_{ab}
	=& {3\over 2}
	\int \frac{\mathrm{d} \mathbf{n}}{4\pi}\;\mathcal{F}_{ab}(\mathbf{n})
	\nonumber\\
	&+ {3\over 8\pi} \int_{z_{\min}}^{z_{\max}} \mathrm{d} z\; c(z)\int \mathrm{d} \mathbf{n}\;\delta_{\rm gal}(\mathbf{n},z)\,\mathcal{F}_{ab}(\mathbf{n})
	\nonumber\\
	=& \Gamma^{\rm HD}_{ab} + 
	{3\over 8\pi}
	\int_{z_{\min}}^{z_{\max}} \mathrm{d} z\; c(z)\,\Gamma^{\delta_{\rm gal}}_{ab}(z;{\bm p}_a, {\bm p}_b), 
	\label{eq:Gamma_iso_plus_lss_z}
\end{align}
where 
\begin{align}
	\Gamma^{\delta_{\rm gal}}_{ab}(z;{\bm p}_a, {\bm p}_b)\equiv \int \mathrm{d} \mathbf{n}\;\delta_{\rm gal}(\mathbf{n},z)\,\mathcal{F}_{ab}(\mathbf{n})
	\label{eq:Gamma_delta_gal_def_main_z}
\end{align}
is the correlation template obtained by weighting the PTA response kernel by the galaxy overdensity map.
The first term in \eqref{eq:Gamma_iso_plus_lss_z} gives the HD ORF term, and the second term gives the anisotropic ORF, i.e., the LSS ORF term tracing the galaxy overdensity map.

We adopt the complex spherical-harmonic convention
\begin{align}
	\delta_{\rm gal}(\mathbf{n}, z)\equiv \sum_{\ell=\ell_{\min}}^{\ell_{\max}}
	\sum_{m=-\ell}^{\ell} a_{\ell m} (z) \,Y_{\ell m}(\mathbf{n}), 
	\label{eq:lss_template}
\end{align}
where 
\begin{align}
	a_{\ell m}(z)=\int \mathrm{d} \mathbf{n}\;\delta_{\rm gal}(\mathbf{n}, z)\,Y^*_{\ell m}(\mathbf{n}).
\end{align}
Define the spherical-harmonic response basis matrices
\begin{align}
	G_{\ell m}^{ab} \equiv \int \mathrm{d}\mathbf{n}\; Y_{\ell m}(\mathbf{n})\,\mathcal{F}_{ab}(\mathbf{n})\,.
	\label{eq:Glm_def}
\end{align}
Then, we have
\begin{align}
	\Gamma^{\delta_{\rm gal}}_{ab} = \sum a_{\ell m} G_{\ell m}^{ab} \,.
\end{align}

\subsection{LSS ORF setting for inference}
\label{subsec:LSS_ORF_setting}

In practice, we discretize the redshift range into $N$ bins,
$[z_{\min},z_{\max}] = \bigcup_{i=1}^{N}[z_i,z_{i+1}]$,
and approximate ${3 c(z) / (8 \pi)}$ as piecewise constant within each bin,
\begin{equation}
	{3 c(z)\over 8 \pi} \equiv \epsilon(z) \simeq \epsilon_i,  \qquad \text{for}\  z\in[z_i,z_{i+1}].
\end{equation}
We then define the binned LSS ORF template as the redshift-integrated contribution in the $i$th bin,
\begin{equation}
	\Gamma^{\rm LSS(i)}_{ab}
	\equiv
	\int_{z_i}^{z_{i+1}} dz\; \Gamma^{\delta_{\rm gal}}_{ab}\!\left(z;{\bm p}_a,{\bm p}_b\right),
	\label{eq:lss_bin_def}
\end{equation}
so that Eq.~(\ref{eq:Gamma_iso_plus_lss_z}) can be written as
\begin{equation}
	\Gamma_{ab}
	\simeq
	\Gamma^{\rm HD}_{ab}
	+ \sum_{i=1}^{N} \epsilon_i\, \Gamma^{\rm LSS(i)}_{ab}\,.
	\label{eq:orf_binned_sum}
\end{equation}

Using the HEALPix \citep[][]{HEALPix2005} pixelization, we discretize the sky into $N_{\rm pix}=12\,N_{\rm side}^2 $, where $N_{\rm side}$ is the HEALPix resolution parameter. In this work, we adopt $N_{\rm side}=64$. Eq.~(\ref{eq:Glm_def}) is then evaluated numerically via a pixel-sum  approximation over pixel centers,
\begin{align}
	G_{\ell m}^{ab} \simeq \Delta\Omega \sum_{i=1}^{N_{\rm pix}}
	Y_{\ell m}(\Omega_i)\,\mathcal{F}_{ab}(\Omega_i)\,,
	\label{eq:Glm_pix}
\end{align}
where $\Delta\Omega \equiv {4\pi/N_{\rm pix}}$ is the solid angle per pixel, and $\Omega_i$ denotes the direction of the $i$th pixel center. Furthermore, finite pixelization can introduce small imaginary components and mild asymmetry; we therefore project the numerically evaluated ORF onto the real, symmetric subspace,
\begin{align}
	\Gamma^{\rm LSS} \leftarrow \Re(\Gamma^{\rm LSS}),\qquad
	\Gamma^{\rm LSS} \leftarrow \frac{1}{2}\left(\Gamma^{\rm LSS}+(\Gamma^{\rm LSS})^{\top}\right). 
\end{align}

\section{Construction of low-$\ell$ LSS spherical-harmonic coefficients}
\label{sec:lss_alm}

In this section we construct low-multipole spherical-harmonic coefficients $a_{\ell m}$ of the LSS galaxy overdensity field for certain redshift slices. Our goal is to obtain a set of stable, cut-sky--robust coefficients up to $\ell_{\max}$ that can be used as input templates in the PTA low-$\ell$ LSS ORF analysis.

\subsection{2MPZ galaxy sample and redshift slicing}
\label{subsec:2mpz}
We use the \emph{2MASS Photometric Redshift} catalog (2MPZ, \cite{Bilicki20142MPZ}) to trace the angular distribution of nearby galaxies and construct redshift-sliced LSS templates for our PTA analysis. 2MPZ contains $\sim 10^6$ galaxies over (nearly) the full sky and provides homogeneous photometric redshifts ($\sigma_z\simeq 0.015$) from cross-matching 2MASS XSC with WISE and SuperCOSMOS \citep[][]{Bilicki20142MPZ}. The catalog primarily probes the low-redshift Universe (with most objects at $z\lesssim 0.3$), making it straightforward to build tomographic templates in the slices $0<z\le0.1$ and $0.1<z\le0.2$; the bin width $\Delta z=0.1$ is substantially larger than the typical photometric-redshift scatter, which helps mitigate inter-bin leakage. 
We restrict the fiducial analysis to these two bins because the available higher-redshift 2MPZ slice is much more sparsely sampled: the $0.2<z\le0.4$ slice contains only $22576$ galaxies before applying the Galactic mask, compared with $630987$ and $279883$ galaxies in the two lower-redshift slices.
This lower tracer density would make the corresponding higher-redshift template more susceptible to shot noise and selection-function systematics.
For the two low-redshift slices considered here, 2MPZ remains well suited for PTA applications: 
its wide sky coverage helps mitigate cut-sky mode coupling for the largest angular modes, its near-/mid-IR selection mitigates sensitivity to Galactic extinction compared to optical-only samples, and its low-redshift depth provides a natural tracer for constructing LSS templates relevant to low-$\ell$ ORF analyses.

We thus divide the catalog into two photometric-redshift bins, $\texttt{bin-1}: 0<z\le0.1$ and $\texttt{bin-2}: 0.1<z\le0.2$, and treat each slice as an independent LSS anisotropy template. From each slice we build overdensity maps on a common pixelization and estimate stable low-$\ell$ spherical-harmonic coefficients $\{a_{\ell m}\}$ up to $\ell_{\max}$ on the masked sky. These coefficients are then used to construct the corresponding PTA LSS-ORF templates, $\Gamma^{\mathrm{LSS}}_{ij}(\texttt{bin-1})$ and $\Gamma^{\mathrm{LSS}}_{ij}(\texttt{bin-2})$, which enter the mixed-ORF models analyzed in the following sections.

\subsection{Galaxy counts and overdensity map in equatorial coordinates}
\label{subsec:overdensity_map}
For each redshift bin we start from a galaxy catalog containing right ascension and declination ($\alpha, \delta$).
We pixelize the catalog into a HEALPix map at resolution $N_{\rm side}$ by counting galaxies per pixel,
\begin{equation}
	N(p) \equiv \sum_{g\in{\rm catalog}} \mathbf{1}\!\left[g\in p\right],
\end{equation}
where $p$ labels HEALPix pixels and $\mathbf{1}$ is the indicator function. 
Throughout, we work in the equatorial (celestial) coordinate system, i.e.\ we identify the spherical angles as
$\theta = \pi/2-\delta$ and $\phi=\alpha$.

We impose a Galactic-plane cut by defining a binary sky mask based on Galactic latitude $b(\mathbf{n})$:
\begin{equation}
	M_{\rm bin}(\mathbf{n})=
	\begin{cases}
		1, & |b(\mathbf{n})|\ge b_{\rm cut},\\
		0, & \text{otherwise},
	\end{cases}
	\label{Galactic_Mask}
\end{equation}
where $b_{\rm cut}=20^\circ$ in our fiducial setup, and $\mathbf{n}$ is the unit vector specifying a sky direction associated with each equatorial HEALPix pixel. Although $b$ is computed in Galactic coordinates, the mask is represented on the same equatorial HEALPix pixelization as the data map.

We define the mean number of galaxies per pixel using the binary mask,
\begin{equation}
	\bar N \equiv 
	\frac{\sum_p M_{\rm bin}(p)\,N(p)}{\sum_p M_{\rm bin}(p)},
\end{equation}
and construct the overdensity field
\begin{equation}
	\delta(p)=
	\begin{cases}
		\displaystyle \frac{N(p)-\bar N}{\bar N}, & M_{\rm bin}(p)=1,\\[6pt]
		0, & M_{\rm bin}(p)=0.
	\end{cases}
	\label{eq:delta_def}
\end{equation}

For subsequent power-spectrum estimation we further apodize the binary mask using a $C^1$ (C1) apodization kernel with scale $\theta_{\rm apod}$,
\begin{equation}
	M_{\rm apod}(\mathbf{n}) \equiv {\rm Apodize}\!\left[M_{\rm bin}(\mathbf{n});\,\theta_{\rm apod},\,{\rm C1}\right],
\end{equation}
with $\theta_{\rm apod}=2^\circ$ in the fiducial setup. The apodized mask is then used as the weight map in the MASTER (pseudo-$C_\ell$) estimation, and is applied consistently in the spherical-harmonic analysis when reconstructing the (masked) harmonic coefficients.

\subsubsection{Low-$\ell$ cleaning: weighted removal of monopole and dipole}
\label{subsubsec:low_l_cleaning}
To mitigate contamination from large-scale offsets and dipolar systematics, and their mode coupling under incomplete sky coverage, we optionally remove the monopole and dipole components of the overdensity field using a weighted least-squares fit on the apodized-mask support.
Such monopole/dipole subtraction is commonly adopted in cut-sky pseudo-$C_\ell$ (MASTER) and correlation-function estimators, where the sky mask couples multipoles and can leak residual large-scale power into higher-$\ell$ modes \citep[e.g.,][]{Hivon2002MASTER,Szapudi2001,Vielva2006ISW,Alonso2019NaMaster}. 

Let $\mathcal{P}=\{p: M_{\rm apod}(p)>\tau\}$ be the set of unmasked pixels after thresholding with $\tau$.
Here $\tau$ is a dimensionless threshold on the apodized mask weight
$M_{\rm apod}(p)$. Since $0\leq M_{\rm apod}(p)\leq 1$, in the fiducial
setup we take $\tau=10^{-3}$ to remove only pixels in the extremely
low-weight boundary region of the apodized Galactic mask. These pixels
carry negligible statistical weight but can mildly degrade the conditioning
of the weighted monopole/dipole fit; the threshold is therefore used solely
as a numerical stability cutoff.
We fit and subtract the $\ell\le1$ subspace,
\begin{equation}
	\delta_{\rm used}(p) \equiv \delta(p) - \sum_{k=1}^{4}\beta_k\,B_k(p),
	\qquad p\in \mathcal{P},
	\label{eq:remove_l01}
\end{equation}
where $\{B_k\}$ are real-valued basis maps spanning the $\ell\le1$ subspace (e.g., $Y_{00}$, $Y_{10}$, and the ${\rm Re}/{\rm Im}$ components of $Y_{11}$) evaluated on the HEALPix grid, and the coefficients $\beta_k$ minimize
\begin{equation}
	{\boldsymbol\beta}=
	\arg\min_{\boldsymbol\beta}\sum_{p\in\mathcal{P}} M_{\rm apod}(p) \Bigl[\delta(p)-\sum_{k=1}^{4}\beta_k B_k(p)\Bigr]^2 \,.
\end{equation}
Pixels outside $\mathcal{P}$ are set to zero in $\delta_{\rm used}$.

\subsection{MAP reconstruction of $a_{\ell m}$ on the cut sky}
\label{subsec:MAP_reconstruction_alm}

We reconstruct the low-$\ell$ harmonic coefficients by fitting a truncated spherical-harmonic expansion to the cut-sky field,
\begin{equation}
	\delta_{\rm used}(\mathbf{n}) \approx 
	\sum_{\ell=2}^{\ell_{\max}}\sum_{m=-\ell}^{\ell} a_{\ell m}\,Y_{\ell m}(\mathbf{n})\,,
	\label{eq:alm_model}
\end{equation}
following the standard Wiener/MAP reconstruction framework for Gaussian random fields with incomplete sky coverage  \citep[e.g.,][]{Lahav1994AllSkyWF,Zaroubi1995WF_Formalism}.
Restricting to pixels $p\in\mathcal{P}$, we form the data vector $y_p\equiv \delta_{\rm used}(p)$ and adopt pixel weights $w_p\equiv M_{\rm apod}(p)$.
To work entirely in real arithmetic, we parameterize the unknown coefficients as
\begin{equation}
	\boldsymbol\theta \equiv \Bigl(\{a_{\ell 0}\},\{\mathrm{Re}\,a_{\ell m}\}_{m>0},\{\mathrm{Im}\,a_{\ell m}\}_{m>0}\Bigr),
\end{equation}
and write the linear model $\mathbf{y}\approx \mathbf{X}\boldsymbol\theta$, where the design matrix $\mathbf{X}$ is constructed by evaluating the corresponding real-valued basis maps on the HEALPix grid.

Assuming Gaussian pixel noise with diagonal weight matrix $\mathbf{W}={\rm diag}(w_p)$, the negative log-likelihood is
\begin{equation}
	\label{eq:likelihood_theta}
	\ln \mathcal{L}(\boldsymbol{\theta}) =
	-{1\over 2}\,(\mathbf{y}-\mathbf{X}\boldsymbol{\theta})^{\top}
	\mathbf{W}
	(\mathbf{y}-\mathbf{X}\boldsymbol{\theta}).
\end{equation}
We further impose a Gaussian prior on the low-$\ell$ harmonic coefficients, with a diagonal prior precision given by the inverse signal covariance as in standard Wiener-filter/MAP reconstructions \citep[e.g.,][]{Lahav1994AllSkyWF,Zaroubi1995WF_Formalism,Bunn1994COBE_WF}. 
In particular, we adopt a data-driven prior covariance by setting its angular power spectrum to the target spectrum $C_\ell^{\rm target}$, estimated from the masked map via MASTER bandpowers computed with \texttt{NaMaster} \citep{Hivon2002MASTER,Alonso2019NaMaster}; see subsection~\ref{subsubsec:Target_spectrum}. 
For a real field this corresponds to $\mathrm{Var}(a_{\ell0})=C_\ell^{\rm target}$ and
$\mathrm{Var}(\mathrm{Re}\,a_{\ell m})=\mathrm{Var}(\mathrm{Im}\,a_{\ell m})=C_\ell^{\rm target}/2$ for $m>0$.
Equivalently,
\begin{subequations}
	\begin{align}
		& \ln \pi(\boldsymbol\theta)
		= -{1\over 2} \,\boldsymbol\theta^{\top}
		\boldsymbol\Lambda\,\boldsymbol\theta,
		\\
		&\Lambda_{kk}=
		\begin{cases}
			1/C_\ell^{\rm target}, & a_{\ell0},\\
			2/C_\ell^{\rm target}, & \mathrm{Re/Im}\,a_{\ell m}\ (m>0).
		\end{cases}
	\end{align}
	\label{eq:prior_theta}
\end{subequations}
where each diagonal element is associated with the appropriate $(\ell,m)$ component.

The MAP estimate ${\boldsymbol\theta}$ via finding the minimal of $(\mathbf{y}-\mathbf{X}\boldsymbol\theta)^{\top}\mathbf{W}(\mathbf{y}-\mathbf{X}\boldsymbol\theta) + \lambda\ \boldsymbol\theta^{\top}\boldsymbol\Lambda\,\boldsymbol\theta$, where $\lambda$ is a tunable regularization strength, scaled in practice to yield numerically stable conditioning. We adopt $\lambda=1$ for the fiducial analysis (see Appendix \ref{app:choice_lambda} for the details). 
Then the solution writes
\begin{equation}
	\left(\mathbf{X}^{\top}\mathbf{W}\mathbf{X}+\lambda\,\boldsymbol\Lambda\right){\boldsymbol\theta}
	=
	\mathbf{X}^{\top}\mathbf{W}\mathbf{y}.
	\label{eq:map_solution}
\end{equation}

The mapping from $\boldsymbol\theta$ back to the complex coefficients $a_{\ell m}$ for $m\ge 0$ is straightforward, and the coefficients for negative $m$ follow from the reality condition
\begin{equation}
	a_{\ell,-m}=(-1)^m\,a_{\ell m}^{\ast}.
	\label{eq:reality}
\end{equation}

\subsubsection{Target spectrum from NaMaster}
\label{subsubsec:Target_spectrum}
We estimate the target angular power spectrum using the MASTER (pseudo-$C_\ell$) approach implemented in \texttt{NaMaster} \citep[][]{Hivon2002MASTER, Alonso2019NaMaster}. 
We define a masked (apodized) field
\begin{equation}
	f(\mathbf{n}) \equiv M_{\rm apod}(\mathbf{n})\,\delta_{\rm used}(\mathbf{n}),
\end{equation}
and compute the coupled pseudo-spectrum $\tilde C_\ell$ from $f$.
In the MASTER formalism, the sky mask couples multipoles such that, in expectation,
\begin{equation}
	\langle \tilde C_\ell \rangle = \sum_{\ell'} \mathbf{M}_{\ell\ell'}\, C_{\ell'} + \tilde N_\ell \,,
\end{equation}
where $\mathbf{M}$ is the mode-coupling matrix determined by the mask, and $\tilde N_\ell$ includes the shot-noise contribution.

We perform a binned (bandpower) deconvolution with linear bins of width $\Delta \ell \equiv n_{\ell b}$,
obtaining deconvolved bandpowers $C_b^{\rm MASTER}$ and their effective multipoles $\ell_{\rm eff}$.
Optionally, we subtract a Poisson shot-noise bias for the number-count overdensity field,
approximated as a white spectrum,
\begin{equation}
	C_{\rm shot}\simeq \frac{1}{\bar n}\,, \
	\bar n \equiv \frac{N_{\rm gal}^{\rm (mask)}}{4\pi f_{\rm sky}}\,, \
	f_{\rm sky}\equiv \frac{1}{4\pi}\int \mathrm{d} \mathbf{n}\, M_{\rm bin}(\mathbf{n}),
\end{equation}
where $N_{\rm gal}^{\rm (mask)}$ is the number of galaxies within the binary mask $M_{\rm bin}$.
Since the spectrum is used only as a positive-definite target prior, we set
$C_b^{\rm MASTER}\leftarrow \max(C_b^{\rm MASTER}-C_{\rm shot},\,0)$.

For use as a prior in the harmonic reconstruction we interpolate the binned spectrum onto the integer multipole grid,
\begin{equation}
	C_\ell^{\rm target} \equiv {\rm interp}\!\left(\ell;\{\ell_{\rm eff}\},\{C_b^{\rm MASTER}\}\right),
	\label{eq:cl_target}
\end{equation}
and enforce a small positive floor to avoid numerical issues at very low power.

\subsubsection{Per-multipole rescaling}
\label{subsubsec:Cl_rescaling}
The MAP solution in Eq.~\eqref{eq:map_solution} yields a set of spherical-harmonic coefficients $a_{\ell m}^{\rm MAP}$ that are numerically stable on the cut sky. 
However, the corresponding full-sky spectrum,
\begin{equation}
	C_\ell^{\rm MAP}\equiv \frac{1}{2\ell+1}\sum_{m=-\ell}^{\ell}\bigl|a_{\ell m}^{\rm MAP}\bigr|^2,
\end{equation}
does not, in general, coincide exactly with the adopted target spectrum $C_\ell^{\rm target}$.
To enforce consistency at the bandpower level, we apply an isotropic, per-multipole rescaling, to the final coefficients $a_{\ell m}$: 
\begin{equation}
	a_{\ell m} = s_\ell\, a_{\ell m}^{\rm MAP},
	\qquad
	s_\ell \equiv \left(\frac{C_\ell^{\rm target}}{C_\ell^{\rm MAP}}\right)^{1/2},
	\label{eq:rescale}
\end{equation}
which guarantees the final spectrum $C_\ell=C_\ell^{\rm target}$ by construction while preserving the relative phases among the $m$-modes at fixed $\ell$. 
The final coefficients $\{a_{\ell m}\}$ for $\ell\in[2,\ell_{\max}]$ are then stored (for both redshift bins) and used to build the corresponding low-$\ell$ LSS ORF templates in the PTA analysis.

\subsection{Low-$\ell$ reconstruction of LSS spherical-harmonic maps}
\label{subsec:lss_alm_maps}

The present analysis utilizes the 15-year NANOGrav timing data for 67 pulsars, determined by observational availability and data quality. Due to the finite and sparse sky coverage of the current array, it is primarily sensitive to the largest angular scales of the SGWB.

For this PTA configuration, the ORFs act as a strong low-pass filter in harmonic space, efficiently suppressing contributions from higher-order multipoles. Consequently, we truncate the LSS harmonic expansion at $\ell_{\max}=12$, which yields a stable PTA-projected LSS ORF template $\Gamma^{\rm LSS}_{ab}$ and is consistent with existing PTA anisotropy studies, which are most sensitive to the lowest angular multipoles \citep[][]{NANOGrav_2023_Anisotropy, Hotinli_2019, Taylor_Gair_2013}.
We verify this choice through ORF-level and timing-likelihood-level
stability tests; see Appendix~\ref{app:ellmax_stability} for details.
Sensitivity to higher-$\ell$ modes is expected to improve with future datasets featuring a larger number of well-distributed pulsars.

To assess whether the recovered low-$\ell$ structure is dominated by counting noise, we compare $C_{\ell}$ reconstructed following the method described in Sections~\ref{subsec:2mpz}--\ref{subsec:MAP_reconstruction_alm} to the overdensity shot-noise level, defined as
\begin{equation}
	\left[\frac{C_\ell}{C_{\rm shot}}\right]_{\ell\in[2,12]}\,.
	\label{eq:median_cl_over_cshot}
\end{equation}
The clustering power at $\ell\in[2,12]$ exceeds the Poisson level by large factors with $C_\ell/C_{\rm shot} \sim \mathcal{O}(10^2)$ for $z<0.1$ and $\sim \mathcal{O}(10^1\text{--}10^2)$ for $0.1<z<0.2$, implying that the recovered low-$\ell$ modes are not dominated by counting noise.

Furthermore, for our fiducial mask choice 
($b_{ \rm cut }=20^\circ$, apodization $2^\circ$), the off-diagonal correlation of the recovered coefficients remains moderate with $\max_{i\neq j}\lvert \mathrm{Corr}(a_i,a_j)\rvert \sim 0.3$ for $\ell\in[2,12]$ (see Appendix \ref{app:offdiag_corr} for the definition), indicating that mode coupling from the cut sky does not lead to catastrophic degeneracies at the multipoles used in this work. These diagnostics support the use of the reconstructed coefficients as a stable low-dimensional summary of the observed LSS for building the ORF templates used in this work.

\begin{figure}[t]
	\centering
	\includegraphics[width=0.48\textwidth]{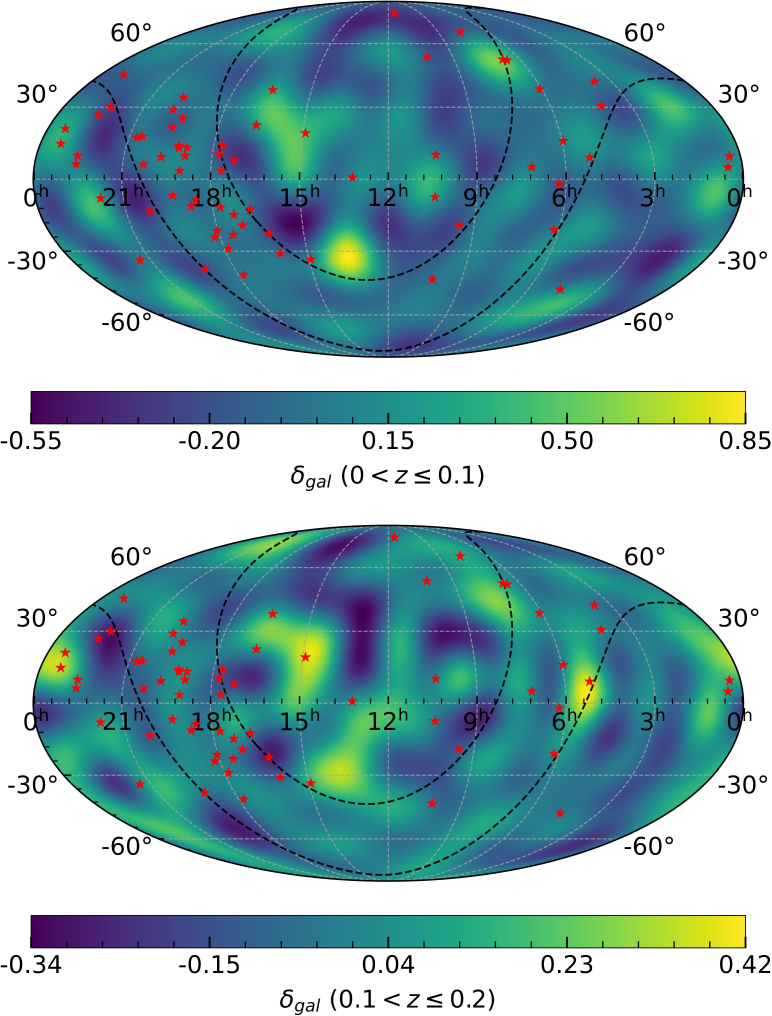}	
	\caption{
		Reconstructed large-scale-structure overdensity maps from spherical-harmonic coefficients $a_{\ell m}$ inferred from the 2MPZ galaxy catalog, truncated to $2 \le \ell \le 12$, for the two redshift bins used in this work. The maps are shown in a Mollweide projection in equatorial coordinates. Red stars mark the sky positions of the 67 pulsars in our PTA sample. The black dashed curves delineate the mask boundary around the Galactic plane, defined using a Galactic latitude cut.
	}
	\label{fig:lss_alm_maps}
\end{figure}

Figure~\ref{fig:lss_alm_maps} displays the reconstructed LSS overdensity maps for two redshift bins from the 2MPZ catalog, following the method outlined in the preceding sections. These maps are derived by truncating the spherical-harmonic expansion of the galaxy overdensity field at $2 \le \ell \le 12$, retaining only the large-angular-scale modes. The black dashed lines delineate the regions corresponding to the Galactic plane mask (Eq.~(\ref{Galactic_Mask})), applied to mitigate stellar contamination and extinction effects.

Within the unmasked sky, the reconstructed maps capture the low-order angular features of the large-scale distribution of nearby galaxies. Differences between the two redshift bins arise from the evolution of LSS and cosmic variance, though similar structures are observed across both maps. These low-$\ell$ maps form the basis for constructing the low-redshift LSS-induced anisotropic ORFs used in the PTA analysis. 

\section{Constraints on LSS-modulated ORFs}
\label{sec:results_lss_orf}

Following the NANOGrav 15-year stochastic-background analysis setup
\citep[][]{NANOGrav_2023_SGWB_evidence,NANOGrav_2024_methods},
we fix the white-noise parameters to the released backend-dependent
values in the 15yr white-noise dictionary.
Timing-model uncertainties are treated with the
\texttt{MarginalizingTimingModel} implementation in
\texttt{enterprise} \citep[][]{enterprise_ref}, as adopted in the public
NANOGrav stochastic-background analysis pipeline
\citep[][]{NANOGrav_2024_methods}. In this approach, the timing-model design
matrix is included as a linear basis in the residual model, and the
corresponding linearized timing-model parameter offsets are analytically
marginalized rather than explicitly sampled. We set
\texttt{use\_svd=True}, as in the public pipeline, to use an
SVD-stabilized timing-model basis for this marginalization, improving
numerical conditioning.
We represent red processes in a Fourier basis with
$f_k=k/T_{\rm span}$, where
$T_{\rm span}=\max(t_{\max})-\min(t_{\min})$ is the array timespan.
We use 30 frequency components for intrinsic pulsar red noise and
$N_f=14$ components for the common power-law SGWB term. We use the same
frequency basis for the ORF-modulated common correlations considered in
this work, matching the NANOGrav 15-year power-law GWB search configuration.

In all cases, the baseline isotropic model corresponds to the HD ORF, while anisotropy is introduced through one or more fixed-shape LSS templates constructed from the 2MPZ galaxy catalog in the low-redshift slices described in Section~\ref{sec:lss_alm}, with each LSS-correlated contribution controlled by a dimensionless mixing coefficient $\epsilon$ (and $\epsilon=0$ reducing to the pure-HD hypothesis).
For the LSS-mixing amplitudes, we adopt independent uniform priors $ \epsilon_i \sim \mathcal{U}(-6,6)$.
This prior is symmetric about zero, allowing both positive and negative correlations with the fixed LSS templates. The same prior range is used for the single-bin and two-bin LSS-modulated ORF models and for the corresponding hypermodel comparisons.

\begin{figure}[t]
	\centering
	\includegraphics[width=0.5\textwidth]{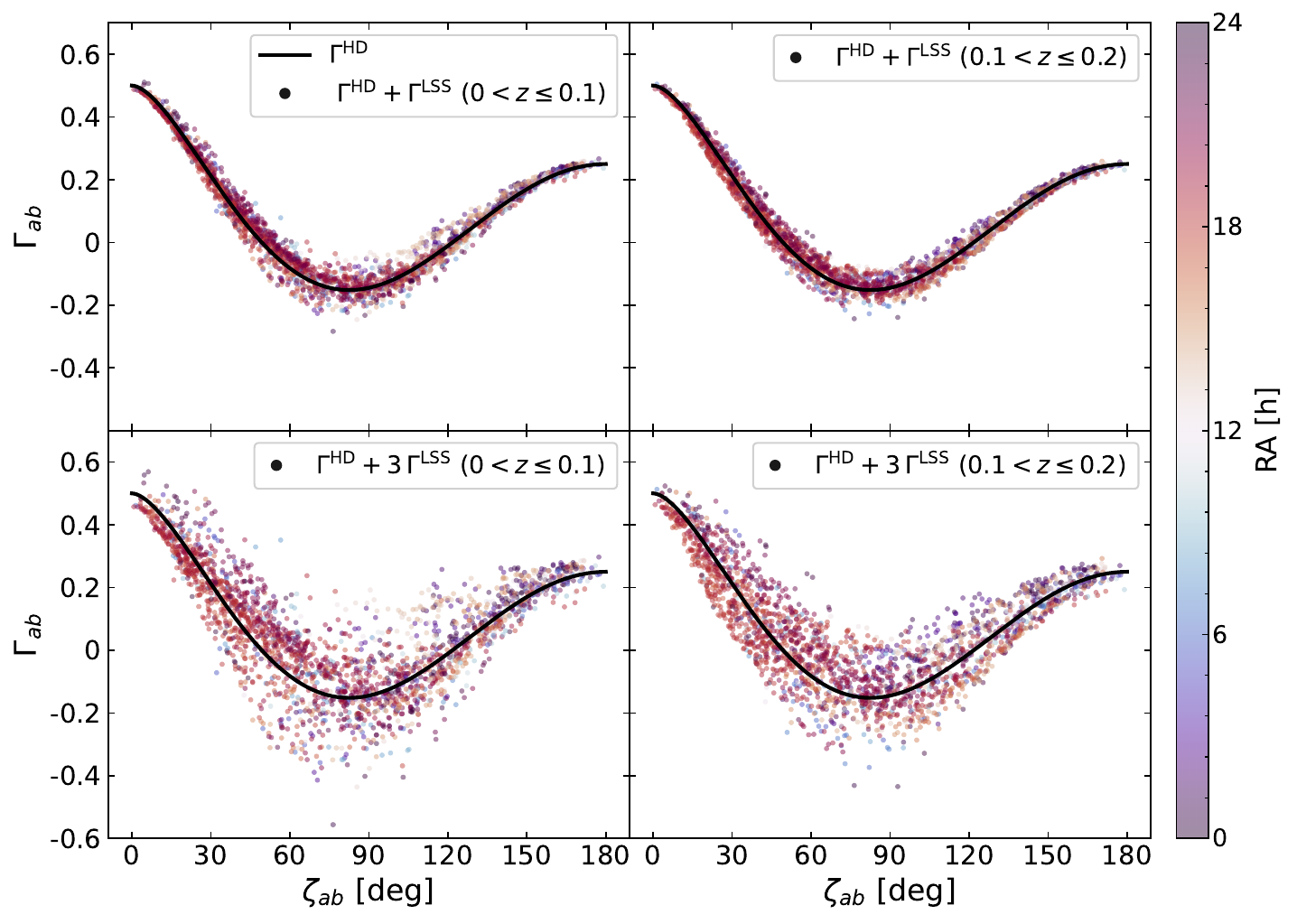}
	\caption{
		Pairwise ORF values $\Gamma_{ab}$ versus pulsar angular separation $\zeta_{ab}$ for the model
		$\Gamma_{ab}=\Gamma^{\rm HD}_{ab}+\epsilon\,\Gamma^{\rm LSS}_{ab}$.
		The solid black curve shows the HD prediction.
		Colored points show $\Gamma_{ab}$ evaluated for individual pulsar pairs and are colored by the pair-midpoint right ascension (RA, in hours).
		Left/right columns correspond to the 2MPZ redshift slices $0<z\le0.1$ and $0.1<z\le0.2$, respectively; top/bottom rows adopt $\epsilon=1$ and $\epsilon=3$.
	}
	\label{fig:hd_lss_orf}
\end{figure}

\subsection{Single-bin LSS-correlated ORF}
\label{subsec:single_bin_lss}

We first consider models in which the isotropic HD correlation is augmented by a single LSS template constructed from one redshift slice,
\begin{equation}
	\Gamma_{ab} \;=\; \Gamma^{\rm HD}_{ab} \;+\; \epsilon_i\, \Gamma^{\rm LSS(i)}_{ab},
\end{equation}
where $i=1,2$ labels the redshift bins, $\Gamma^{\rm LSS(i)}_{ab}$ is the corresponding LSS-derived ORF template (Section~\ref{sec:lss_alm}), and $\epsilon_i$ parametrizes the relative amplitude of the LSS-correlated anisotropic component in that bin.
Positive (negative) $\epsilon_i$ corresponds to a component correlated (anti-correlated) with the LSS template.

Figure~\ref{fig:hd_lss_orf} illustrates how an LSS-correlated contribution modifies the HD correlation at the level of individual pulsar pairs.
For an isotropic background, $\Gamma_{ab}$ is uniquely determined by the separation $\zeta_{ab}$, producing a single curve.
Adding the anisotropic template $\Gamma^{\rm LSS}$ breaks this one-to-one mapping: pairs with the same $\zeta_{ab}$ can have different $\Gamma_{ab}$ depending on their sky locations relative to the fixed LSS pattern.
This leads to a characteristic spread of pairwise ORF values around the HD curve, with a magnitude that increases with $|\epsilon|$ (compare $\epsilon=1$ to $\epsilon=3$).
The detailed distribution of points differs between the two redshift slices because the underlying LSS maps are not identical, leading to different projections onto the sparse PTA sky sampling.

\begin{figure}[h]
	\centering
	\includegraphics[width=0.95\columnwidth]{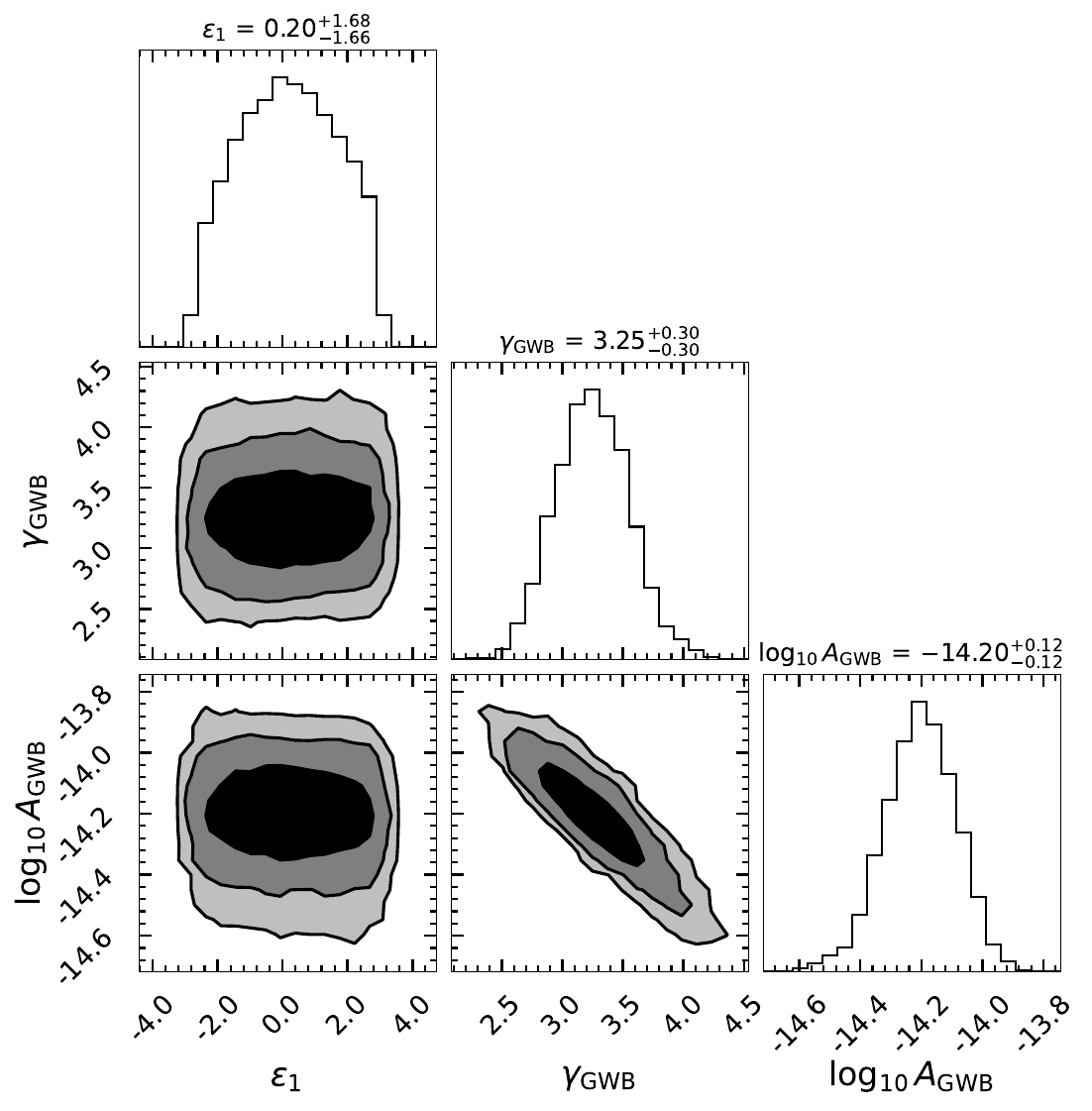}
	\caption{
		Marginalized posterior distribution for $(\epsilon_1,\gamma_{\rm gw},\log_{10}A_{\rm gw})$ in the single-bin model
		$\Gamma=\Gamma^{\rm HD}+\epsilon_1\,\Gamma^{\rm LSS(1)}$, where $\Gamma^{\rm LSS(1)}$ is derived from the 2MPZ slice $0<z\le0.1$.
		Diagonal panels show 1D marginalized posteriors and off-diagonal panels show joint credible regions.
		Titles report the median and $68\%$ credible intervals.
	}
	\label{fig:posteriors_bin1}
\end{figure}

\begin{figure}[h]
	\centering
	\includegraphics[width=0.95\columnwidth]{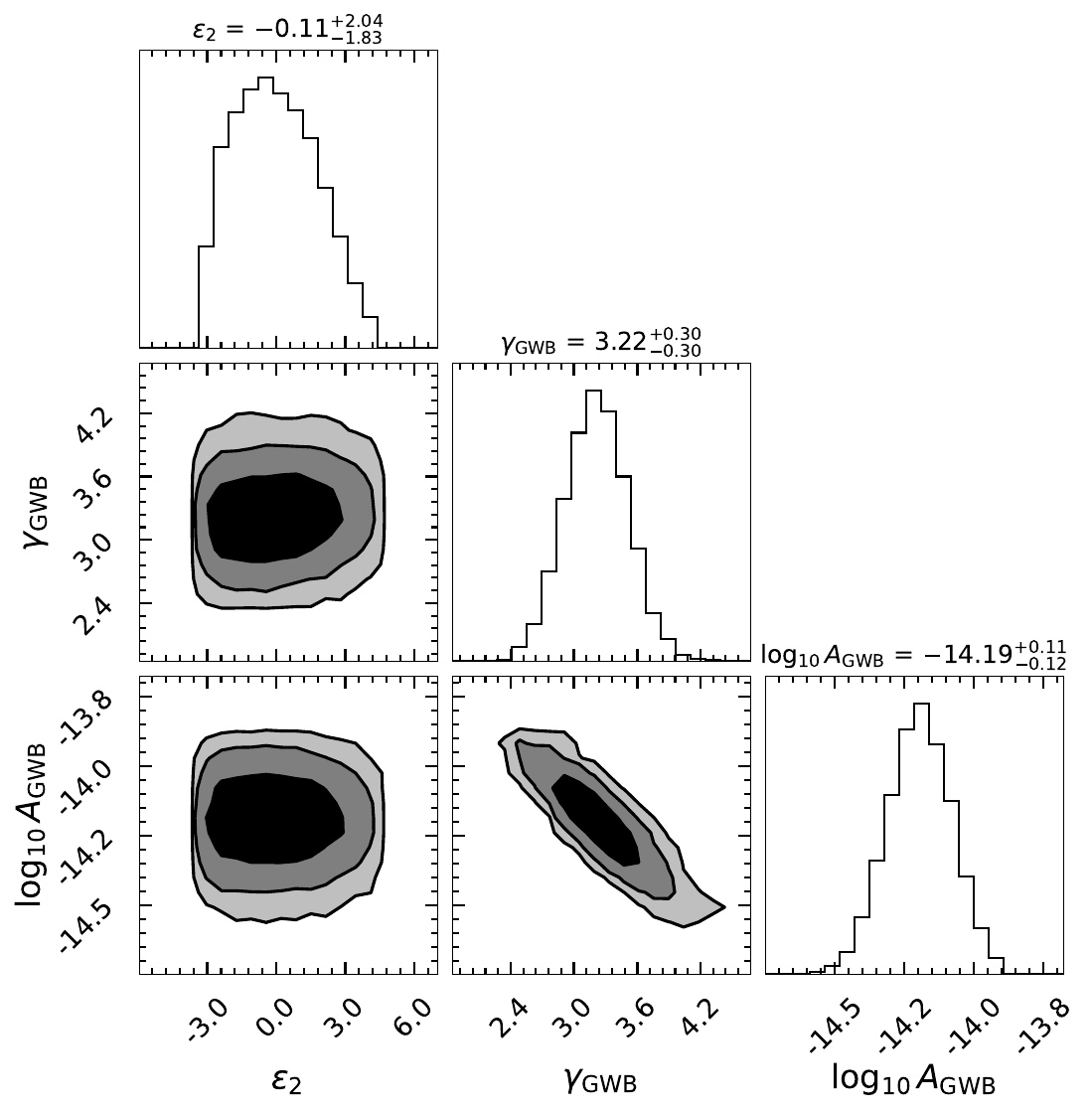}
	\caption{
		Same as Figure~\ref{fig:posteriors_bin1}, but for $(\epsilon_2,\gamma_{\rm gw},\log_{10}A_{\rm gw})$ in the model
		$\Gamma=\Gamma^{\rm HD}+\epsilon_2\,\Gamma^{\rm LSS(2)}$, where $\Gamma^{\rm LSS(2)}$ is derived from the 2MPZ slice $0.1<z\le0.2$.
	}
	\label{fig:posteriors_bin2}
\end{figure}

The posterior distributions for the single-bin mixed-ORF models are shown in Figures~\ref{fig:posteriors_bin1} and \ref{fig:posteriors_bin2}.
In both redshift slices, the inferred mixing amplitude is consistent with $\epsilon_i=0$, with
$\epsilon_1 = 0.20^{+1.68}_{-1.66}$ and $\epsilon_2 = -0.11^{+2.04}_{-1.83}$ (median and $68\%$ credible intervals).
The breadth of these posteriors implies that, within the sensitivity of the present dataset, the analysis primarily yields upper limits on the amplitude of any 2MPZ-template-correlated contribution, rather than a statistically significant detection of correlated structure.
Any detailed, pair-by-pair substructure in the LSS-correlated ORF presented in Figure~\ref{fig:hd_lss_orf} is therefore either absent or not yet statistically resolvable with the current 15-year dataset.
Meanwhile, the GWB spectral parameters $\log_{10}A_{\rm GWB}$ and $\gamma_{\rm GWB}$ retain their usual power-law degeneracy, and their joint posterior shows no pronounced correlation with $\epsilon_i$, indicating that allowing an LSS-modulated degree of freedom does not materially shift the inferred isotropic-background parameters.

\subsection{Two-bin LSS-correlated ORF}
\label{subsec:two_bin_lss}
We next consider a more general phenomenological ORF model that allows independent LSS-correlated contributions from both 2MPZ redshift slices,
\begin{equation}
	\Gamma_{ab}
	\;=\;
	\Gamma^{\rm HD}_{ab}
	\;+\;
	\epsilon_1\,\Gamma^{\rm LSS(1)}_{ab}
	\;+\;
	\epsilon_2\,\Gamma^{\rm LSS(2)}_{ab},
\end{equation}
where $\epsilon_1$ and $\epsilon_2$ parameterize the amplitudes of the two fixed-shape LSS-modulated templates constructed from the low-redshift 2MPZ slices.

\begin{figure}[h]
	\centering
	\includegraphics[width=1.0\columnwidth]{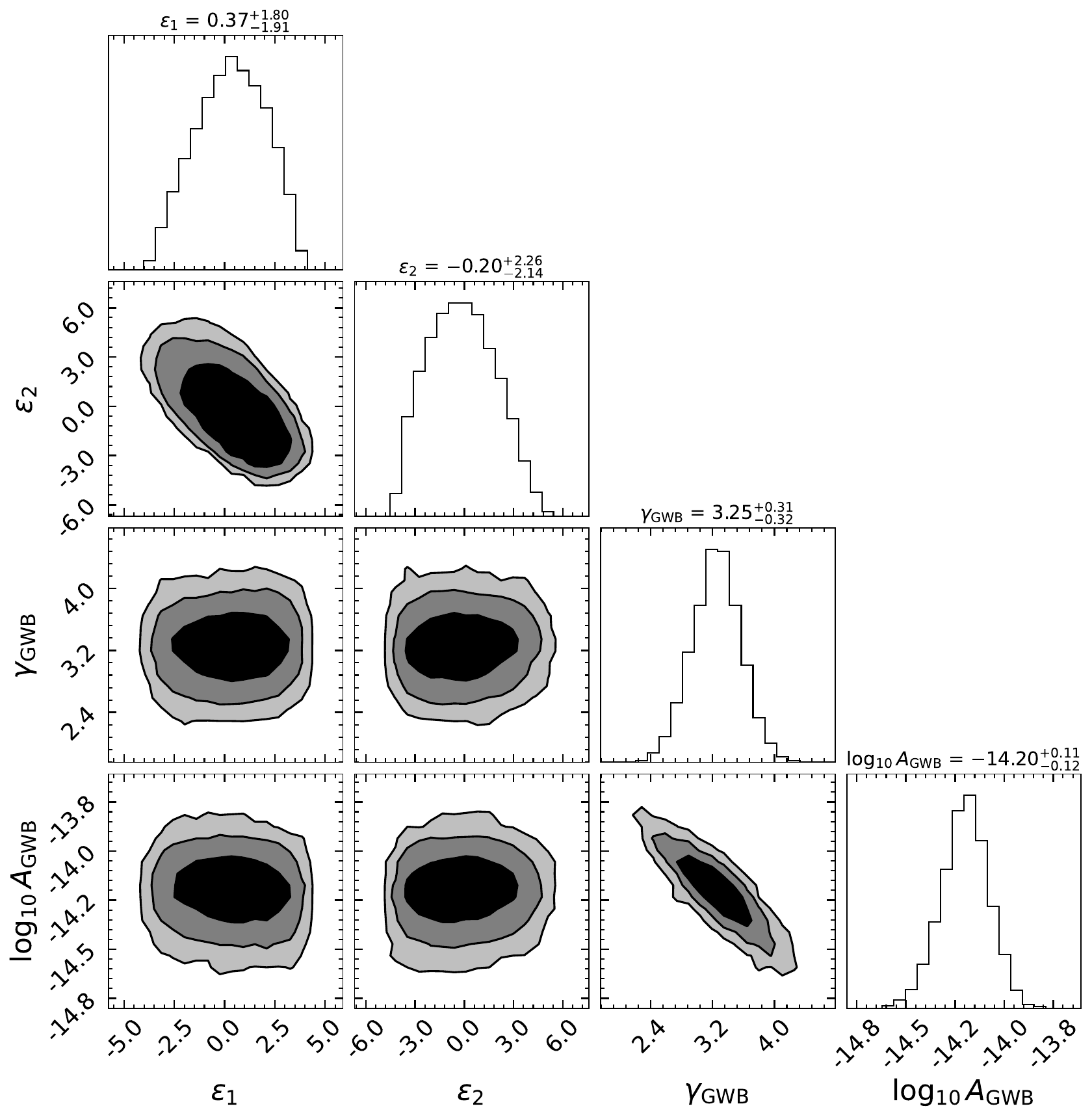}
	\caption{
		Marginalized posterior distribution for $(\epsilon_1,\epsilon_2,\gamma_{\rm gw},\log_{10}A_{\rm gw})$ in the two-bin model
		$\Gamma=\Gamma^{\rm HD}+\epsilon_1\Gamma^{\rm LSS(1)}+\epsilon_2\Gamma^{\rm LSS(2)}$.
		The joint posterior exhibits a strong degeneracy between $\epsilon_1$ and $\epsilon_2$, consistent with partial non-orthogonality of the two LSS templates under the PTA sampling of pulsar pairs.
		Titles report the median and $68\%$ credible intervals.
	}
	\label{fig:posteriors_bin1_bin2}
\end{figure}

The marginalized posterior for the two-bin model is shown in Figure~\ref{fig:posteriors_bin1_bin2}.
Both anisotropy parameters remain consistent with zero,
$\epsilon_1 = 0.37^{+1.80}_{-1.91}$ and $\epsilon_2 = -0.20^{+2.26}_{-2.14}$ (median and $68\%$ credible intervals),
while the inferred $(\log_{10}A_{\rm gw},\gamma_{\rm gw})$ are consistent with the single-bin analyses.
A prominent feature of the joint posterior is an elongated, tilted support in the $(\epsilon_1,\epsilon_2)$ plane, indicating a strong degeneracy between the two coefficients.
This behavior reflects the fact that the current 15-year PTA dataset does not cleanly disentangle two non-orthogonal, fixed-shape low-redshift LSS templates when projected onto the finite set of pulsar pairs; instead, it primarily constrains particular linear combinations of $(\epsilon_1,\epsilon_2)$.

\begin{figure}[h]
	\centering
	\includegraphics[width=1.0\columnwidth]{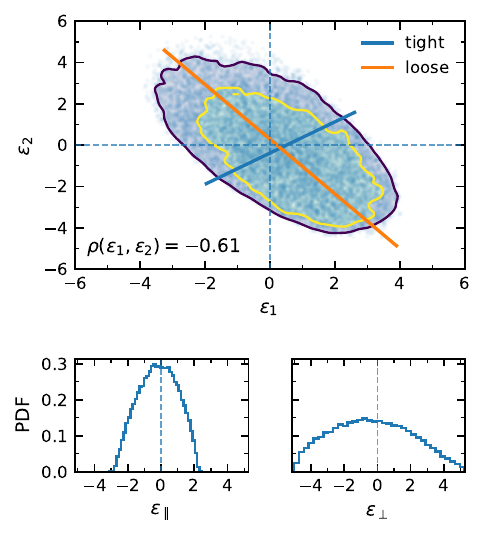} 
	\caption{
		PCA visualization of the $(\epsilon_1,\epsilon_2)$ degeneracy in the two-bin model.
		\emph{Top:} joint posterior in the $(\epsilon_1,\epsilon_2)$ plane with 68\% and 95\% credible contours, overlaid with the principal axes of the posterior covariance (``tight'' and ``loose'').
		The annotation reports the Pearson correlation coefficient $\rho(\epsilon_1,\epsilon_2)$ (Eq.~(\ref{Pearson_coefficient})).
		\emph{Bottom:} marginalized posteriors of the projected combinations $\epsilon_{\parallel}$ (tightly constrained principal component) and $\epsilon_{\perp}$ (more weakly constrained orthogonal component), illustrating that the data predominantly constrain one linear combination of the two LSS template amplitudes.
	}
	\label{fig:eps_pca_bin1_bin2}
\end{figure}

To quantify and visualize this degeneracy, we compute the Pearson correlation coefficient
\begin{equation}
	\rho(\epsilon_1,\epsilon_2)\equiv
	\frac{{\rm Cov}(\epsilon_1,\epsilon_2)}
	{\sqrt{{\rm Var}(\epsilon_1)\,{\rm Var}(\epsilon_2)}},
	\label{Pearson_coefficient}
\end{equation}
which is $\rho(\epsilon_1,\epsilon_2)=-0.61$ for this run, indicating a substantial anti-correlation.
We further rotate $(\epsilon_1,\epsilon_2)$ into the principal-component basis of their posterior covariance matrix.
Concretely, letting $\mathbf{C}$ denote the $2\times2$ covariance of $(\epsilon_1,\epsilon_2)$, we diagonalize $\mathbf{C}$ and use its orthonormal eigenvectors to define new coordinates
\begin{equation}
	\begin{pmatrix}
		\epsilon_{\parallel}\\
		\epsilon_{\perp}
	\end{pmatrix}
	=
	\mathbf{U}^{\top}
	\begin{pmatrix}
		\epsilon_1\\
		\epsilon_2
	\end{pmatrix},
\end{equation}
where $\mathbf{U}$ is the orthonormal matrix whose columns are the eigenvectors of $\mathbf{C}$.
By construction, $\epsilon_{\parallel}$ corresponds to the eigen-direction with the smaller eigenvalue and is therefore the tightly constrained combination, whereas $\epsilon_{\perp}$ corresponds to the larger-eigenvalue direction and is the more weakly constrained orthogonal combination.
Figure~\ref{fig:eps_pca_bin1_bin2} summarizes the Principal Component Analysis (PCA) interpretation: the 68\% and 95\% credible regions in the $(\epsilon_1,\epsilon_2)$ plane are elongated along the principal axes, and the 1D posteriors of $\epsilon_{\parallel}$ and $\epsilon_{\perp}$ highlight the well- and poorly-constrained combinations, respectively.

\subsection{Bayesian model comparison}
\label{subsec:hypermodel_bf}

We evaluate the relative support of the NANOGrav 15-year dataset for an LSS-modulated common-spectrum process, compared to the standard isotropic HD hypothesis, using hypermodel sampling. In this framework, a discrete model indicator is sampled jointly with the continuous timing-noise and GW parameters, enabling a direct inference of the evidence ratios between competing hypotheses.

For the single-bin LSS-correlated ORFs discussed in Section \ref{subsec:single_bin_lss}, we obtain
\begin{equation}
	\mathcal{B}_{{\rm HD+LSS}_1,\,{\rm HD}} = 0.40 \,,
	\qquad
	\mathcal{B}_{{\rm HD+LSS}_2,\,{\rm HD}} = 0.43 \,,
\end{equation}
both below unity. These values indicate that the current data do not provide evidence for an additional LSS-modulated degree of freedom beyond the isotropic HD correlations. Accordingly, the single-bin analyses constrain the amplitude of any SGWB component whose angular structure follows the low-redshift galaxy distribution traced by 2MPZ, while remaining consistent with statistical isotropy.

Finally, when allowing for a combined low-redshift LSS contribution, as has been discussed in Section \ref{subsec:two_bin_lss}, we find
\begin{equation}
	\mathcal{B}_{{\rm HD+LSS_1+LSS_2},\,{\rm HD}} = 0.11\,.
\end{equation}
The further reduction in the Bayes factor reflects the increased model complexity associated with introducing two LSS-correlation amplitudes: with the present PTA sensitivity and sky sampling, the data do not warrant the additional freedom, and the simpler isotropic HD model is preferred.

\section{Discussion and Conclusion}
\label{sec:discu_and_concl}
In this work, we present the first Bayesian PTA likelihood analysis that embeds an externally observed, full-sky galaxy-survey LSS template directly as an ORF component. 
In our template-based framework, the pulsar-pair cross-correlations are modeled as a linear mixture of the isotropic Hellings--Downs (HD) form and one or more fixed, tomographic LSS-correlated templates,
\begin{equation}
	\Gamma_{ab}=\Gamma^{\rm HD}_{ab}+\sum_i \epsilon_i \Gamma^{\rm LSS(i)}_{ab}\,,
\end{equation}
where $\Gamma^{\rm LSS(i)}_{ab}$ is constructed from the 2MPZ galaxy catalog \citep[][]{Bilicki20142MPZ} in redshift bin $i$. Here we consider two tomographic bins, $0<z\leq0.1$ and $0.1<z\leq0.2$. 
For a chosen normalization of $\Gamma^{\rm LSS(i)}_{ab}$, the coefficient $\epsilon_i$ quantifies the relative admixture of an SGWB component whose pairwise angular correlations project onto that fixed LSS template, while the common-spectrum GW amplitude is fitted simultaneously with timing-noise parameters. 

Applying this framework to the NANOGrav 15-year dataset \citep[][]{NANOGrav_2023_SGWB_evidence,NANOGrav_dataset_2023}, we find no evidence for an LSS-correlated SGWB component traced by the 2MPZ templates. The posteriors are consistent with $\epsilon_i=0$ in both single-bin and two-bin analyses (Figures~\ref{fig:posteriors_bin1}--\ref{fig:posteriors_bin1_bin2}), and the hypermodel Bayes factors are all below unity
($\mathcal{B}_{{\rm HD+LSS}_1,\,{\rm HD}}=0.40$,
$\mathcal{B}_{{\rm HD+LSS}_2,\,{\rm HD}}=0.43$,
$\mathcal{B}_{{\rm HD+LSS}_{1+2},\,{\rm HD}}=0.11$),
indicating that the present data do not justify introducing additional LSS-correlated degrees of freedom beyond the HD hypothesis. This outcome is broadly consistent with existing PTA anisotropy searches, which mainly constrain departures from isotropy in the lowest-order multipoles beyond the monopole and currently report upper limits rather than detections of anisotropy \citep[e.g.,][]{Taylor_2015_PRL,NANOGrav_2023_Anisotropy,MPTA_2024_Anisotropy,PPTA_2026_Anisotropy,Sah2026}.

An important feature of the inference is that introducing the LSS-modulated amplitudes $\epsilon_i$ does not noticeably alter the recovered isotropic-background parameters: the posteriors show no pronounced correlation between $\epsilon_i$ and the GWB spectral parameters $\log_{10}A_{\rm GWB}$ and $\gamma_{\rm GWB}$ (Figures~\ref{fig:posteriors_bin1}--\ref{fig:posteriors_bin1_bin2}).
By contrast, the two-bin LSS ORF model exhibits a pronounced degeneracy between $(\epsilon_1,\epsilon_2)$, with $\rho(\epsilon_1,\epsilon_2)=-0.61$ (Figures~\ref{fig:posteriors_bin1_bin2}--\ref{fig:eps_pca_bin1_bin2}). 
This behavior is expected: the two low-$z$ templates are partially non-orthogonal in the low multipoles to which PTAs are most sensitive, and sparse/non-uniform pulsar sky sampling further reduces the effective dimensionality of the response.
The PCA rotation (Figure~\ref{fig:eps_pca_bin1_bin2}) indicates that only one linear combination is preferentially constrained, whereas the orthogonal combination remains largely prior-dominated.
Consequently, increasing the number of redshift slices without additional information (e.g., improved PTA angular resolution, orthogonalized template bases, or external priors) may primarily increase degeneracies rather than substantially tighten per-bin constraints. 
For the present dataset, the two-bin analysis should therefore be interpreted primarily as an exploratory tomographic extension and a diagnostic of the effective dimensionality of the accessible low-redshift template space, rather than as evidence that two redshift-separated LSS-correlated components are already independently resolved.

It is also important to emphasize that the inferred amplitudes
$\epsilon_i$ have a template-dependent meaning. They quantify the
component of the PTA pairwise correlations that is aligned with the
adopted low-redshift LSS ORF templates, rather than providing a
model-independent measure of SGWB anisotropy. 
Consequently, the constraints on $\epsilon_i$ should not be interpreted as generic bounds on the total anisotropic power of the SGWB, or as a direct measurement of the fraction of the expected SGWB anisotropy signal power arising from $z<0.2$.

In principle, a perfect detection could expect a full cross-correlation between the SGWB and the galaxy field which tells the information of the relation between these two across angular scale, redshift, and potentially frequency. However, carrying out such an analysis generally requires a sufficiently reliable recovery of the GW anisotropy field from PTA data, or at least a GW map estimate that can be directly cross-correlated with galaxy maps. 
Given the current number of pulsars in PTAs, limited sky coverage, and modest signal-to-noise ratio for anisotropy, this remains challenging for present datasets. 
Previous work has suggested that a statistically significant detection of the SGWB--galaxy cross-correlation will likely become substantially more feasible only with future PTA datasets, especially in the SKA era \citep[][]{sah2024,Cusin_2025_CrossCorre}. 
A major practical advantage of the LSS ORF template approach is that it does not require reconstruction of the full GW sky. Instead, it tests directly within the PTA timing likelihood for a specific, physically motivated correlation pattern. This makes it a more data-native strategy, better suited to PTA datasets with low signal-to-noise ratios and limited angular resolution for anisotropy. 

This compression, however, also makes the inference sensitive to the adopted ORF template and to how much of the true SGWB anisotropy is coherent with that template. 
In particular, realization-dependent ``shot-noise'' anisotropy can arise from Poisson fluctuations in source counts and sky locations when the nanohertz background is produced by a finite number of contributing binaries rather than by a perfectly smooth Gaussian field \citep[e.g.,][]{Ravi_2012,Cornish_2013,Allen2024_PTA_anisotropy,Taylor_2020PRD,Becsy_2022,Konstandin_2024}. Such a component is expected to be only weakly correlated with the underlying LSS structure. In practice, it acts as additional angular structure that increases variance, broadens the posterior on $\epsilon_i$, and can suppress evidence for LSS modulation even if the ensemble-averaged SGWB is astrophysical and traces the matter distribution in a statistical sense. 

Our fiducial ansatz further assumes that the LSS-correlated component of the SGWB anisotropy is a maximally correlated, phase-aligned copy of the
observed galaxy overdensity field. 
Equivalently, for $c(z)>0$, their angular cross-correlation coefficient satisfies
$C_\ell^{\rm gw\times gal}/
{\sqrt{C_\ell^{\rm gw}C_\ell^{\rm gal}}}=1$.
In general, however, the SGWB anisotropy physically associated with the LSS need not be a perfectly phase-aligned copy of the observed galaxy field, so that
$C_\ell^{\rm gw\times gal}/
{\sqrt{C_\ell^{\rm gw} C_\ell^{\rm gal}}}<1$.
Such a reduction can arise from differences between GW-luminosity weighting and galaxy-number weighting, stochasticity in the SMBH--galaxy--halo connection, mismatch between the SGWB and galaxy redshift kernels, large-scale survey systematics, or additional SGWB components that are not correlated with the LSS. A reduced correlation coefficient would decrease the fraction of the anisotropic signal that projects coherently onto the fixed
2MPZ-derived ORF template, thereby weakening the constraint on the total anisotropic contribution and reducing the Bayesian support for the HD+LSS model.

Within these methodological and data limitations, the present PTA data mainly provide upper limits on the amplitude of any 2MPZ-template-correlated contribution.
Any detailed, pair-by-pair substructure in the ORF associated with such an LSS-correlated component (Figure~\ref{fig:hd_lss_orf}) is therefore either absent or not yet statistically resolvable with the current 15-year dataset. 
This non-detection suggests that a robust probe of such correlated structure will require improved PTA sensitivity, a denser and more uniform pulsar sky distribution, together with wider-area, higher-number-density, and better-characterized galaxy surveys. 

Overall, this work provides a first dedicated search for LSS-correlated anisotropy using real PTA data within a Bayesian ORF-template likelihood framework. 
Looking ahead, two developments are especially well-motivated by the limitations identified here: (i) systematic injection-and-recovery studies to forecast detectability and expected constraints on $\epsilon_i$ as a function of PTA sensitivity and pulsar sky coverage, and (ii) extension of the LSS-correlated ORF framework to higher redshifts using multiple wide-area galaxy surveys.
These steps will turn the present analysis into a predictive program: not only constraining LSS-correlated anisotropy in current data, but also forecasting when and how future PTAs can detect, or place tight bounds on, an SGWB component that traces the cosmic web. We leave these studies for future work. 

\begin{acknowledgments}
This work was supported by the National Natural Science Foundation of China (Grant No.~12405068). 
This work used publicly released NANOGrav 15-year data products \citep[][]{NANOGrav_dataset_2023,NANOGrav_dataset_2025}; the analysis and interpretations are those of the author and do not necessarily reflect the views of the NANOGrav Collaboration. 
We thank Siyuan Chen, Qing-Guo Huang, Zu-Cheng Chen, and Huiyuan Wang for useful discussions.   
This work made use of the 2MASS Photometric Redshift (2MPZ) galaxy catalog \citep[][]{Bilicki20142MPZ} and open-source Python packages, including \texttt{enterprise} \citep[][]{enterprise_ref}, \texttt{enterprise\_extensions} \citep[][]{enterprise_extensions_ref}, \texttt{PTMCMCSampler} \citep[][]{PTMCMCSampler_ref}, \texttt{NaMaster} \citep[][]{Hivon2002MASTER, Alonso2019NaMaster}, \texttt{HEALPix}\citep[][]{HEALPix2005}, and \texttt{healpy} \cite[][]{healpy2019}. Numerical calculations were performed on the HPC cluster of the National Supercomputing Center in Zhejiang.
\end{acknowledgments}

%

\appendix
\section{Diagnostics for the low-$\ell$ reconstruction and LSS-template stability}
\label{app:lowell_diagnostics}
	This appendix summarizes three diagnostics used in the main text to assess
	the robustness of the low-$\ell$ reconstruction and the resulting LSS ORF
	templates. First, we quantify the maximum absolute off-diagonal correlation
	among the recovered spherical-harmonic coefficients, which diagnoses mode
	coupling induced by the cut sky. Second, we describe the choice of the
	regularization strength $\lambda$ adopted in the MAP reconstruction of the
	spherical-harmonic coefficients
	(Section~\ref{subsec:MAP_reconstruction_alm}). Third, we test the stability
	of the PTA-projected LSS ORF templates and of the corresponding timing
	likelihood with respect to the harmonic truncation $\ell_{\max}$.
	
	
	\subsection{Off-diagonal correlation of recovered coefficients}
	\label{app:offdiag_corr}
	The likelihood and prior described in Eqs.~(\ref{eq:likelihood_theta}) and~(\ref{eq:prior_theta}) yield the coefficient covariance
	\begin{equation}
		\mathrm{Cov}(\boldsymbol\theta)\;\approx\; (\mathbf{X}^{\top} \mathbf{W} \mathbf{X}+\boldsymbol\Lambda)^{-1}\,,
		\label{eq:theta_cov}
	\end{equation}
	which propagates the measurement uncertainty to the recovered coefficients. 
	Mode coupling induced by the cut sky and weighting can introduce correlations among the recovered coefficients. We quantify this through the correlation matrix
	\begin{equation}
		\mathrm{Corr}(\theta_i, \theta_j)\;\equiv\;\frac{\mathrm{Cov}_{ij}}{\sqrt{\mathrm{Cov}_{ii}\,\mathrm{Cov}_{jj}}},
		\label{eq:corr_mat}
	\end{equation}
	constructed from the covariance in Eq.~\eqref{eq:theta_cov}. As a compact scalar diagnostic of inter-mode coupling we report
	\begin{equation}
		\max_{i\neq j}\,|\mathrm{Corr}(\theta_i, \theta_j)|, \!
		\quad \text{or equivalently} \quad \!
		\max_{i\neq j}\lvert \mathrm{Corr}(a_i,a_j)\rvert \,, 
		\label{eq:max_offdiag_corr}
	\end{equation}
	i.e., the maximum absolute off-diagonal correlation among all distinct parameter pairs. Values $\ll 1$ indicate weak coupling between modes, while values approaching unity would signal near-degeneracies caused by mask-induced mixing.
	
	\subsection{Choice of the regularization strength $\lambda$}
	\label{app:choice_lambda}
	The parameter $\lambda$ controls the bias--variance trade-off of the cut-sky inversion in Eq.~\eqref{eq:map_solution}.
	Since the reconstructed coefficients are ultimately used only through the induced PTA overlap-reduction function (ORF),
	we select $\lambda$ by requiring the ORF to be stable against further increases in regularization.
	For each trial $\lambda$ we construct the corresponding ORF matrix $\Gamma_{ij}(\lambda)$ and quantify the change between
	neighboring values using the off-diagonal relative Frobenius difference,
	\begin{equation}
		r(\lambda_a,\lambda_b)\equiv
		\left[
		\frac{\sum_{i<j}\big(\Gamma_{ij}(\lambda_a)-\Gamma_{ij}(\lambda_b)\big)^2}
		{\sum_{i<j}\Gamma_{ij}^2(\lambda_b)}
		\right]^{1/2},
	\end{equation}
	together with the off-diagonal Pearson correlation coefficient $\rho(\lambda_a,\lambda_b)$.
	We then choose the smallest $\lambda$ lying on the stability plateau, i.e. satisfying
	$r(\lambda,\lambda_{\rm next})\ll 1$ and $\rho(\lambda,\lambda_{\rm next})\simeq 1$ for subsequent $\lambda$ values.
	In our scan we find that $\lambda\simeq 1$ is the onset of this plateau; we therefore adopt $\lambda=1$ for the fiducial analysis.
	We verified that varying $\lambda$ within the plateau (e.g. $\lambda\in[1,2]$) leaves the ORF essentially unchanged.

	\subsection{Stability with Respect to the Harmonic Truncation}
	\label{app:ellmax_stability}
	In the main text we adopt a truncated spherical-harmonic representation of the galaxy overdensity field when constructing the LSS-correlated ORF template. Here we clarify the ``convergence'' test of our fiducial choice of $\ell_{\max}=12$. It is important to emphasize that ``convergence'' here refers to the effective stability of the PTA-projected ORF, and of the timing likelihood built from it, under further increases of the harmonic truncation. 
	
	Throughout this appendix, we denote by
	\begin{equation}
		\Gamma_{ab}^{\rm LSS}[L]
		\equiv
		\sum_{\ell=2}^{L}\sum_{m=-\ell}^{\ell}
		a_{\ell m}\,G_{\ell m}^{ab}
		\label{eq:app_glss_L}
	\end{equation}
	the PTA-projected LSS ORF synthesized with harmonic truncation $\ell_{\max}=L$. Since the diagonal entries are fixed by convention, the convergence tests below are performed on the off-diagonal pulsar-pair correlations only.
	
	To compare two truncations $L_1<L_2$, we vectorize the upper-triangular off-diagonal entries of the ORF matrix,
	\begin{equation}
		\mathbf{v}^{(L)} \equiv \left\{\Gamma_{ij}^{\rm LSS}[L]\right\}_{i<j},
	\end{equation}
	and define the relative Frobenius difference
	\begin{equation}
		r(L_1,L_2)=\left[\frac{\sum_{k}\left(\mathbf{v}_{k}^{(L_2)}-\mathbf{v}_{k}^{(L_1)}\right)^2}{\sum_{k}\left(\mathbf{v}_{k}^{(L_2)}\right)^2}\right]^{1/2}
	\end{equation}
	together with the Pearson correlation coefficient
	\begin{equation}
		\mathrm{Corr}(L_1,L_2)
		\equiv
		\frac{
			\sum_k
			\left(v_k^{(L_1)}-\bar v^{(L_1)}\right)
			\left(v_k^{(L_2)}-\bar v^{(L_2)}\right)
		}{
			\sqrt{\sum_k \left(v_k^{(L_1)}-\bar v^{(L_1)}\right)^2}
			\sqrt{\sum_k \left(v_k^{(L_2)}-\bar v^{(L_2)}\right)^2}
		},
		\label{eq:app_corr_def}
	\end{equation}
	where $k$ labels pulsar pairs and $\bar v^{(L)}$ denotes the mean of the vectorized off-diagonal entries. The quantity $r$ measures the overall relative change in the off-diagonal ORF entries, while $\mathrm{Corr}$ quantifies the similarity of the pulsar-pair correlation pattern independent of an overall rescaling.
	
	Using the sequence $L=3,6,12,30$, we find for the bin-1 template
	\begin{equation}
		r(3,30)=0.76 \,, \,\,\,
		r(6,30)=0.35 \,, \,\,\,
		r(12,30)=0.13,
	\end{equation}
	with corresponding correlation coefficients
	\begin{equation}
		\mathrm{Corr}(3,30)=0.69\,, \,\,
		\mathrm{Corr}(6,30)=0.94\,, \,\,
		\mathrm{Corr}(12,30)=0.99.
	\end{equation}
	For the bin-2 template, we obtain
	\begin{equation}
		r(3,30)=0.61 \,, \,\,\,
		r(6,30)=0.33 \,, \,\,\,
		r(12,30)=0.14,
	\end{equation}
	with
	\begin{equation}
		\mathrm{Corr}(3,30)=0.79 \,, \,\,
		\mathrm{Corr}(6,30)=0.94 \,, \,\,
		\mathrm{Corr}(12,30)=0.99.
	\end{equation}
	These results indicate that the differences between the truncated ORF templates decrease as the maximum multipole is increased. In particular, the fiducial choice $\ell_{\max}=12$ already captures the main shape of the binned ORF templates, with the remaining total-amplitude difference being below approximately $15\%$ relative to the $L=30$ result.
	
	Since current PTA data are primarily sensitive to low angular multipoles, increasing the truncation beyond, for example, $\ell_{\max}=12$ is expected to increase the computational cost without adding significant information. This expectation can be verified directly by comparing the timing likelihoods obtained with $\ell_{\max}=12$ and $\ell_{\max}=30$ over posterior-supported samples. Specifically, we define
	\begin{equation}
		\Delta \ln \mathcal{L}_s(12,30)
		\equiv
		\ln \mathcal{L}_s[30]-\ln \mathcal{L}_s[12],
		\label{eq:app_dlnL_def}
	\end{equation}
	where $s$ labels a posterior sample, and $\ln \mathcal{L}_s[L]$ denotes the log-likelihood of that sample evaluated using the LSS ORF template $\Gamma_{ab}^{\rm LSS}[L]$. In this test, we randomly select 200 posterior-supported samples from the corresponding chains, each including the full set of 137 model parameters.
	
	For the bin-1 template, we find
	\begin{equation}
		\Delta\ln\mathcal{L}(12,30)
		=
		+0.0067^{+0.088}_{-0.052}\,,
	\end{equation}
	where the uncertainties denote the 16th--84th percentile range. For the bin-2 template, we obtain
	\begin{equation}
		\Delta\ln\mathcal{L}(12,30)
		=
		-0.015^{+0.276}_{-0.303}\,.
	\end{equation}
	Thus, increasing the truncation from $\ell_{\max}=12$ to $\ell_{\max}=30$ produces a negligible change in the timing likelihood for both redshift bins.
	
	Taken together, the ORF-level and timing-likelihood tests suggest that the current PTA acts as a strong low-pass filter in angular space, with the observable pulsar-pair correlation structure being primarily sensitive to the lowest multipoles. In our analysis, a fiducial truncation of $\ell_{\max}=12$ is sufficient to capture the relevant low-multipole structure of the LSS ORF templates, while remaining a conservative choice for the likelihood analysis. Increasing the truncation beyond this value introduces only minor corrections and does not qualitatively change either the pairwise correlation pattern or the timing likelihood.

\bibliography{references}{}
\bibliographystyle{aasjournalv7.1}

\end{document}